\documentclass[slantedGreek,fleqn,submission,copyright,creativecommons]{eptcs} 

\providecommand{\event}{WLP'15} 
\title{A Practical View on Renaming} 

\author{Marija Kula\v{s}
\institute{FernUniversit\"{a}t in Hagen, Wissensbasierte Systeme, 58084 Hagen, Germany} 
\email{kulas.marija@online.de}
}
\def\titlerunning{A Practical View on Renaming}
\def\authorrunning{M.~Kula\v{s}}

\hypersetup{%
  pdftitle={A Practical View on Renaming},
  pdfauthor={Marija Kulas}
}

\usepackage{breakurl} 
\usepackage{amsmath}
\usepackage{amssymb} 
\usepackage{amscd} 
\usepackage{stmaryrd} 
\usepackage{array}
\usepackage{xspace} 
\usepackage{graphicx} 
\usepackage[dvipsnames,usenames]{color}
\usepackage{centernot} 
\usepackage{enumitem} 
\setlist{itemsep=2pt, parsep=2pt} 
\usepackage{booktabs} 
\usepackage{amsthm, thmtools}

\newcommand{\reference}[1]{}  

\newenvironment{cell}[1][]
{\begin{tabular}[t]{@{}>{#1}l@{}}}%
{\end{tabular}}

\newenvironment{myproof}{\begin{proof}}{\end{proof}}

\declaretheorem[style=definition, numberwithin=section]{definition}
\declaretheorem[style=definition, sibling=definition]{axiom}
\declaretheorem[style=definition, sibling=definition]{example}

\declaretheorem[style=plain, sibling=definition]{theorem}
\declaretheorem[sibling=definition]{lemma}
\declaretheorem[sibling=definition]{legacy} 
\declaretheorem[sibling=definition]{corollary}

\declaretheorem[style=remark, sibling=definition]{remark}
\declaretheorem[style=remark, sibling=definition]{notation}

\makeatletter
\newcommand{\ifempty}[1]{%
    \if\relax\detokenize{#1}\relax%
        \expandafter\@firstoftwo
    \else
        \expandafter\@secondoftwo
    \fi}
\makeatother

\definecolor{darkblue}{rgb}{0,0,0.4}
\def\Commentcolor{darkblue} 
\def\coloremph{blue}
\def\coloremphsym{\coloremph}

\newcommand{\myemph}[1]{{\it\color{\coloremph} #1}}  
\newcommand{\myemphsym}[1]{{\color{\coloremphsym} #1}}  

\newcommand{\mymodelold}{S1:PP\xspace} 

\newcommand{\yt}{this author\xspace} 

\newcommand{\any}{\_} 

\newcommand{\ie}{i.e.\@}

\makeatletter
\newcommand*{\defeq}{\mathrel{\rlap{%
                     \raisebox{0.3ex}{$\m@th\cdot$}}%
                     \raisebox{-0.3ex}{$\m@th\cdot$}}%
                     =}
\makeatother

\newcommand{\sei}{\ensuremath{\defeq}\xspace}

\newcommand{\suchthat}{\mathrel{\mid}} 

\newcommand{\meta}[1]{\ensuremath{\mathit{#1}}} 

\newcommand{\fsig}[2][]{\ifempty{#1}{\meta{#2}}{\ensuremath{\meta{#2}/#1}}}

\newcommand*{\from}{\colon} 

\newcommand{\goto}{\ensuremath{\rightsquigarrow}} 

\newcommand{\bfsl}[1]{\text{\bfseries\slshape#1}\xspace}

\newcommand{\Nat}{\bfsl{N}} 
\newcommand{\VarSet}{\bfsl{V}} 
\newcommand{\FunctorSet}{\bfsl{Fun}}

\newcommand{\presek}{\cap} 
\newcommand{\unija}{\cup} 
\newcommand{\dunija}{\uplus} 

\newcommand{\mytop}[2]{\genfrac{}{}{0pt}{}{#1}{#2}} 

\newcommand{\idsubst}{\ensuremath{\varepsilon}\xspace} 

\newcommand{\eqsub}[1][]{\ensuremath{\sim_{#1}}} 

\newcommand{\compos}{\ensuremath{\cdot}} 

\newcommand{\ko}[1]{\centernot{#1}} 

\newcommand{\vardis}{\ensuremath{\ko{\bowtie}}} 
\newcommand{\nonvardis}{\ensuremath{\bowtie}}  

\newcommand{\id}[1]{\ensuremath{\varepsilon_{#1}}} 

\newcommand{\restrict}[2]{\ensuremath{#1\mathord\upharpoonright_{#2}}} 

\newcommand{\variant}[1][]{\ensuremath{\ifempty{#1}{\cong}{=_{#1}}}}  

\newcommand{\realDom}{\meta{Dom}\xspace} 
\newcommand{\realRange}{\meta{Range}\xspace} 

\newcommand{\varsname}{\mathit{Vars}}  
\newcommand{\corename}{\mathit{Core}}  
\newcommand{\cocorename}{\mathit{Ran}} 

\newcommand{\varsof}[1]{\ensuremath{\varsname({#1})}}  
\newcommand{\Dom}{\ensuremath{\corename}}  
\newcommand{\Range}{\ensuremath{\cocorename}} 

\newcommand{\coreplusname}{\ensuremath{C}}  
\newcommand{\cocoreplusname}{\ensuremath{R}}  
\newcommand{\varsplusname}{\ensuremath{V}}  

\newcommand{\vplus}[1]{\ensuremath{\varsplusname(#1)}\xspace} 
\newcommand{\cplus}[1]{\ifempty{#1}{\coreplusname}{\ensuremath{\coreplusname(#1)}}\xspace} 
\newcommand{\rplus}[1]{\ifempty{#1}{\cocoreplusname}{\ensuremath{\cocoreplusname(#1)}}\xspace} 

\newcommand{\sadd}{\dunija} 

\newcommand{\indom}{\meta{InDom}\xspace} 

\newcommand{\noninj}{\meta{Pit}\xspace} 

\newcommand{\closure}[1]{\ensuremath{\overline{#1}}} 

\newcommand{\pren}{\ensuremath{\mathit{Pren}}\xspace} 

\newcommand{\unrelax}[1]{{\ensuremath{[#1]}}}

\newcommand{\trenam}{prenaming\xspace} 
\newcommand{\Trenam}{Prenaming\xspace} 

\newcommand{\cumulative}{relevant\xspace} 
\newcommand{\Cumulative}{Relevant\xspace} 
\newcommand{\cumulativeness}{relevance\xspace}

\newcommand{\clausi}{\ensuremath{\mathcal{K}}\xspace}
\newcommand{\klausi}{\bar{\mathcal{K}}} 

\newcommand{\emptycl}{\ensuremath{\boxempty}\xspace}  

\newcommand{\gneck}{\ensuremath{\leftarrow}\xspace}

\newcommand{\compd}{\bfsl{D}}  

\newcommand{\cas}{c.a.s.\@} 
\newcommand{\Cas}{C.a.s.\@} 

\newcommand{\casq}{complete answer\xspace} 

\newcommand{\dopp}{\mathord{:}}

\newcommand{\resymbolup}{\ensuremath{\lhook\joinrel\relbar\joinrel\mkern1mu\mathrel{\raise.1ex\hbox{\text{\footnotesize{$\triangleright$}}}}}}

\newcommand{\resol}[1]{\ensuremath{\resymbolup_{#1}}}  

\newcommand{\specu}{\bfsl{X}} 
\newcommand{\specv}{\bfsl{Y}} 

\newcommand{\mgus}{\ensuremath{\mathit{Mgus}}\xspace} 

\newcommand{\ualg}{\bfsl{U}}  

\newcommand{\hclparam}{{\text{HCL}(\ualg)}} 

\begin{document}
\maketitle

\begin{abstract}

We revisit variable renaming from a practitioner's point of view,
presenting concepts we found useful in dealing with operational semantics of pure Prolog.
A concept of \emph{relaxed core representation} is introduced, 
upon which a concept of \emph{prenaming} is built. 
Prenaming formalizes the intuitive practice of renaming terms by 
just considering the necessary bindings,
where now some passive "bindings" x/x may be necessary as well.
As an application, a constructive version of variant lemma for
implemented Horn clause logic has been obtained.
There, prenamings made it possible to incrementally handle new
(\emph{local}) variables.

\end{abstract}

\section{Introduction}

For logic program analysis or formal semantics, the issue of 
renaming variables and generally handling substitutions is inevitable. 
Yet the image of substitutions in logic programming research is a
somewhat tainted one, at least since it has been pointed out by H.-P. Ko
\cite[p.\,148]{shep:94} that the original claim of strong completeness
of SLD-resolution needs to be amended, 
due to a counter-example using the fact that \(\left(\mytop{x}{f(y,z)}\right)\)
is not more general than \(\left(\mytop{x}{f(a,a)}\right)\). 
The example may look counter-intuitive, but it complies with the
definition of substitution generality. 
Also, by composing substitutions, properties like equivalence, idempotency 
or restriction are not preserved.  
Lastly, due to group structure of renamings, permuting any number of variables 
amounts to "doing nothing", as in \(\left(\mytop{x}{y} \mytop{y}{x}\right) \eqsub \idsubst\), 
and such equivalences are also felt to be counter-intuitive.  
Hence the prevalent sentiments that substitutions are "a quite hard matter
to deal with" (\cite{catuscia:90}) or "very tricky" (\cite{shep:94}).
As a remedy, in the context of aggregating most general unifiers in a
logic programming computation some helpful new concepts and operators were proposed, 
like \emph{parallel composition} instead of traditional composition (\cite{catuscia:90}) 
and \emph{resultant} instead of answer substitution (\cite{lloyd:shepherdson:91}).
Still, for almost anyone embarking on a journey of logic program
analysis or formal semantics, sooner or later the need for renaming variables
and generally handling substitutions in a new context arises.

In case of \yt, the need arose while trying to prove adequacy of
an operational semantics for pure Prolog, \mymodelold \cite{KulasM:towcb}, 
and the context was one of \emph{extensibility}: 
Given is a pair of queries that are alphabetic variants of each other. 
As their respective \mymodelold derivations proceed to develop,
new variables may crop up, due to standardization-apart 
(here called \emph{local variables}, \autoref{sec:local}),  
but the status of being variant should hold.
This setup is known from the classical \emph{variant lemma} (\cite{lloyd}).  
Additionally, the corresponding variables need to be \emph{collected},
obtaining at each step the temporary variance between the derivations.
As an example, assume the first query is \(p(z,u,x)\) and the second \(p(y,z,x)\).  
There is only one relevant renaming, 
\(\rho=\left(\mytop{z}{y} \mytop{u}{z} \mytop{y}{u}\right)\).  
Now assume in the next step the first derivation acquires the variable \(y\), 
and the second \(w\).  
The relevant renaming this time would be 
\(\rho'=\left(\mytop{z}{y} \mytop{u}{z} \mytop{y}{w} \mytop{w}{u}\right)\).  
Clearly, \(\rho'\) is not an extension of \(\rho\),
which makes it seem unsafe to proceed: are some properties of the
previous step now in danger?
So the question is, how to "safely" extend a variable mapping.
For this purpose, in \autoref{sec:trenam} we introduce a slight
generalization of renaming, called \emph{\trenam{}}. 
It is a mathematical underpinning of the intuitive practice of
renaming terms by just considering the necessary bindings, and not
worrying whether the result is a permutation. 
In the above example, 
renaming \(p(z,u,x)\) to \(p(y,z,x)\) means mapping
\(z\mapsto y,\, u\mapsto z\) and \( x\mapsto x\). 
Intuitively, only \(z\mapsto y,\, u\mapsto z\) are considered necessary bindings,
giving the "renaming" \(\left(\mytop{z}{y} \mytop{u}{z}\right)\).
For \trenam, \(x\mapsto x\) is necessary as well.
It is based on \emph{relaxed core representation}, which is nothing else than
allowing some \(x\mapsto x\) pairs alongside "real" bindings, as placeholders.

\Trenam{}s relate to and are inspired by previous work as follows. 
In \cite{plotkin:phd}, the concept of \emph{translation} is defined,
upon which alphabetic variance and standardization apart are built;
this is the same as \trenam but for relaxed core (page \pageref{page-transl}). 
A \emph{safe} \trenam is more general than \emph{renaming for a term}
from \cite{lloyd}, and it maximizes \(W\) in the notion of
\emph{W-renaming} from \cite{eder:85} (page \pageref{page-w-ren}).  
Also, it generalizes \emph{substitution renaming} 
from \cite{amato:05} (\autoref{sec:substvar}).

In \autoref{sec:appl}, \trenam{}s are used to express and prove
a propagation claim for implemented Horn clause logic,
by means of local variable extension (\autoref{lem:propag}).
As a corollary, a variant lemma is obtained (\autoref{lem:var}).  
Underway, we touch on the discrepancy between
the rather abundant theory of logic programming and a scarcity of
mathematical claims for implemented logic programming systems.  While
there are some formal proofs of properties like nominal unification
\cite{urban-unif}, for logic programming systems or their compilation
such are still few and far between, a notable exception being \cite{pusch-wam}.  
New concepts like \trenam may be of help.

\section{Substitution}

First we need a bit of notation. 
Assume two disjoint sets: 
a countably infinite set \myemphsym{\VarSet} of \myemph{variable}s 
and a set \myemphsym{\FunctorSet} of \myemph{shape}s.
If \(W \subseteq \VarSet\), any mapping \(F\) with \(F(W)\subseteq\VarSet\) shall be called
\myemph{variable-pure on} \(W\).
A mapping variable-pure on the whole set of variables \VarSet shall be 
simply called \myemph{variable-pure}. 
If \(\VarSet\setminus W\) is finite, \(W\) is said to be \myemph{co-finite}.
A mapping \(F\) is \myemph{injective on} \(W\),
if whenever \(F(x)=F(y)\) for \(x,y \in W\) also holds \(x=y\).
Each 
\(\myemphsym{\fsig[n]{f}} \in \FunctorSet\) consists of
a \myemph{functor} \(f\) and an associated number of arguments \(n\),
called \myemph{arity}.
Functors of arity \(0\) are called \myemph{constant}s.  Starting from
\VarSet and \FunctorSet, data objects or \myemph{term}s%
\footnote{In Prolog, everything is a term, and so shall \emph{term} be here 
the topmost syntactic concept.} are built:
Any variable $x \in \VarSet$ is a {term}.
If \(t_1,...,t_n\) are terms and \(\fsig[n]{f} \in \FunctorSet\),
then \(f(t_1,...,t_n)\) is a {term} with \myemph{shape} \fsig[n]{f}
and \myemph{constructor} \(f\).  In case of \(\fsig[0]{f}\), the
term shall be written without parentheses.
If a term \(s\) occurs within a term \(t\), we write \myemphsym{\(s \in t\)}.
The \myemph{ordered pair} of terms \(h\) and \(t\) is written 
as \myemphsym{\([h|t]\)},
where \(h\) is called the \emph{head} and \(t\) the \emph{tail} of the
pair.  A special case is a \myemph{non-empty list}, 
distinguished by its tail being a special term 
\(\myemphsym{[]}\) called the \myemph{empty list}, or a non-empty list
itself.  A \myemph{list of \(n\) elements} is the term
\([t_1|[t_2|[...[t_n|[]]]]]\), conveniently written as \myemphsym{\([t_1,...,t_n]\)}.
Let \myemphsym{\varsof{t}} be the set of variables in the term $t$.
If the terms \(s\) and \(t\) share a variable, that shall be written
\myemphsym{\(s \nonvardis t\)}.
Otherwise, we say \(s,t\) are \myemph{variable-disjoint}, written as
\myemphsym{\(s \vardis t\)}.

A recurrent theme in this paper shall be "relevance",
meaning "no extraneous variables" relative to some term or terms.  
It was used in \cite[p.\.38]{apt:book} in unary sense, \ie{} no extraneous
variables relative to (one) term.  This usage shall be reflected in the
text as follows: A renaming \(\rho\) embedding a \trenam \(\alpha\) is a
\myemph{relevant} embedding, if \(\varsof{\rho} \subseteq
\varsof{\alpha}\) (\autoref{alg:embed}).  Additionally, relevance in a
binary sense, concerning two terms, shall also be needed: 
\label{page-cumul}
A mapping \(F\) is \myemph{\cumulative for \(t_1\) to \(t_2\)}, 
if \( \realDom(F)\subseteq \varsof{t_1} \) and \(\realRange(F) \subseteq \varsof{t_2}\) 
(\autoref{alg:pren}, \autoref{lem:propag}).

\begin{definition}[substitution]
A \myemph{substitution} $\theta$ is a function mapping variables to terms, 
which is identity almost everywhere. 
In other words, it is a function \(\theta\) with 
domain \(\realDom(\theta)=\VarSet\) such that 
the set $\myemphsym{\Dom(\theta)} \sei \{x \in \VarSet \suchthat \theta(x)\neq x\}$
is finite.%
\footnote{\cite{gallier} speaks of \emph{finite support}.
}

The set \(\Dom(\theta)\) shall be called the \myemph{active domain}%
\footnote{%
Traditionally called just \emph{domain}.
This may be confusing, since in the usual mathematical sense 
it is always the whole \VarSet that is the domain of any substitution.
}
or \myemph{core} of $\theta$, 
and its elements \myemph{active variable}s%
\footnote{The name \emph{active variable} appears in \cite{jacobs:langen:1992}.}
of \(\theta\).
The set \(\myemphsym{\Range(\theta)} \sei \theta(\Dom(\theta))\) 
is the \myemph{active range} of $\theta$.
A variable \(x\) such that $\theta(x) = x$ shall be called 
a \myemph{passive} \emph{variable}, or a \emph{fixpoint}, for \(\theta\).  
Also, we say that \(\theta\) is \emph{active} on the variables from
\(\Dom(\theta)\), and \emph{passive} on all the other variables.
If $\Dom(\theta) = \{x_1,..., x_k\}$, where $x_1,..., x_k$ are pairwise
distinct variables, and $\theta$ maps each $x_i$ to $t_i$, then $\theta$
shall have the \myemph{core representation} \(\{x_1/t_1, ..., x_k/t_k\}\),
or the perhaps more visual 
\(\left(\mytop{x_1}{t_1} \mytop{...}{...} \mytop{x_k}{t_k}\right)\).
Each pair \(x_i,t_i\) is called the \myemph{binding} for $x_i$ in $\theta$,
denoted by \myemphsym{\(x_i/t_i \in \theta\)}.
Often we identify a substitution with its core representation, and thus
regard it as a syntactical object, a term representing a finite set.  So
the set of variables of a substitution is defined as
$\myemphsym{\varsof{\theta}} \sei \Dom(\theta)\unija\varsof{\Range(\theta)}$. 

The notions of restriction and extension of a mapping shall also be
transported to core representation: if \(\theta \subseteq \sigma\), we say
\(\theta\) is a \myemph{restriction} of \(\sigma\),
and 
\(\sigma\) is an \myemph{extension} of \(\theta\).
The restriction \myemphsym{$\restrict{\theta}{W}$}
of a substitution $\theta$ on a set of variables $W\subseteq\VarSet$
is defined as follows:
if $x\in W$ then $\restrict{\theta}{W}(x)\sei \theta(x)$,
otherwise $\restrict{\theta}{W}(x)\sei x$.
The restriction of $\theta$ upon the variables of $t$ is abbreviated as
$\myemphsym{\restrict{\theta}{t}} \sei \restrict{\theta}{\varsof{t}}$.

The \myemph{composition} $\myemphsym{\theta\compos\sigma}$ 
of substitutions $\theta$ and $\sigma$ is defined by 
$(\theta\compos\sigma)(x) \sei \theta(\sigma(x))$.
Composition may be iterated, written as
\(\myemphsym{\sigma^n}\sei\sigma\compos\sigma^{n-1}\) for \(n\geq 1\),
and \(\myemphsym{\sigma^0}\sei\idsubst\).
Here \(\myemphsym{\idsubst} \sei \left(\right)\)  
is the identity function on \VarSet.
In case a variable-pure substitution \(\rho\) is bijective, its inverse
shall be denoted as \(\myemphsym{\rho^{-1}}\).  
A substitution $\theta$ satisfying the equality $\theta\compos\theta =\theta$ 
is called \myemph{idempotent}.

Definition of substitution is enhanced from variables to arbitrary terms
in a structure-preserving way by
$\myemphsym{\theta(f(t_1,..., t_n))} \sei f(\theta(t_1),..., \theta(t_n))$. 
If $t$ is a term, then $\theta(t)$ is an \myemph{instance} of $t$ via $\theta$.
\end{definition} 

\begin{example}
\(
\left(\mytop{x}{u} \mytop{w}{v} \mytop{u}{x} \mytop{v}{w}\right) \compos \left(\mytop{u}{x} \mytop{v}{w} \mytop{x}{y} \mytop{y}{u} \mytop{z}{v} \mytop{w}{z}\right)
= \left(\mytop{\ko{u}}{\ko{u}} \mytop{\ko{v}}{\ko{v}} \mytop{x}{y} \mytop{y}{x} \mytop{z}{w} \mytop{w}{z} \mytop{\ko{x}}{\ko{u}} \mytop{\ko{w}}{\ko{v}} \mytop{\ko{u}}{\ko{x}} \mytop{\ko{v}}{\ko{w}}\right)  
= \left(\mytop{x}{y} \mytop{y}{x} \mytop{z}{w} \mytop{w}{z}\right)
\).
\end{example}

\section{Renaming}

\begin{definition}[renaming]\label{def:ren}
A \myemph{renaming} of variables is a bijective variable-pure substitution.
\end{definition}
\noindent
In \cite{eder:85}, it is synonymously called "permutation". 
We shall reserve the word for the general case where movement of
infinitely many variables is possible.
Here we synonymously speak of \myemph{finite permutation}
due to the fact that, being a substitution, any renaming has a finite
core, and \autoref{lem:permut} holds.

Due to structure preserving, if \(s \in t\) then \(\sigma(s) \in \sigma(t)\).
For bijective substitutions (\ie{} renamings), 
the converse property holds as well, giving

\begin{lemma}[renaming stability of "{$=$}", "{$\in$}", "{$\vardis$}"]\label{lem:ren:aff}  
Let \(\rho\) be a renaming and \(s,t\) be terms.
Then
\(s=t\) ~iff~ \(\rho(s)=\rho(t)\),
and also
\(s\in t\) ~iff~ \(\rho(s) \in \rho(t)\).
As a consequence, \(s\vardis t\) ~iff~ \(\rho(s) \vardis \rho(t)\).
\end{lemma}

\begin{legacy}[\cite{maher}]\label{lem:permut}  
A substitution \(\rho\) is a renaming iff \( \rho(\Dom(\rho)) = \Dom(\rho)\).
\end{legacy}

\begin{legacy}[\cite{eder:85}]\label{lem:eder:85}
Every injective variable-pure substitution is a renaming.
\end{legacy}

So composition of renamings is a renaming.  The next property is about
cycle decomposition of a finite permutation.

\begin{lemma}[cycles]\label{lem:cyc}
Let \(\sigma\) be a variable-pure substitution.  It is injective iff 
for every \(x\in\VarSet \) there is \( n\in\Nat\) such that \(\sigma^n(x)=x\).
\end{lemma}
\begin{myproof}\reference{lem:cyc}
Assume \(\sigma\) injective, and choose \(x_0\in\VarSet\).  
If \(\sigma(x_0)=x_0\), we are done.  Otherwise,
\(\sigma^i(x_0)\not=\sigma^{i-1}(x_0)\) for all \(i\ge 1\), due to
injectivity.  Hence, \(\sigma^{i-1}(x_0) \in \Dom(\sigma)\) for every
\(i\ge 1\).  Because of the finiteness of \(\Dom(\sigma)\), there is \(
m>k\geq 1\) such that \( \sigma^m(x_0)=\sigma^k(x_0) \).  Due to
injectivity, \(\sigma^{m-1}(x_0)=\sigma^{k-1}(x_0)\).  By iteration we
get \(n\sei m-k\).
For the other direction, assume \(\sigma(x)=\sigma(y)\), and minimal
\(m,n\) such that \(\sigma^n(x)=x,~\sigma^m(y)=y\).  Consider the case
\(m\not=n\), say \(m>n\).  
Then \(\sigma^{m-n}(y)=\sigma^{m-n}(x)=\sigma^{m-n}(\sigma^n(x))=
\sigma^{m-n}(\sigma^n(y))= \sigma^{m}(y)=y \), contradicting minimality
of \(m\).  
Hence \(m=n\), so \(x=\sigma^{n}(x)=\sigma^{n}(y)=y\).
\end{myproof}

\section{Relaxed core representation} 

If there is a substitution \(\sigma\) mapping a term \(s\) on a term
\(t\), then it is mapping each variable in \(s\) on a subterm of \(t\).
It is possible that a variable stays the same, 
so if we want our mapping to explicitely cover
\emph{all} variables in \(s\), 
as in the promised application (\autoref{sec:appl}), then
necessarily \(x/x\) would have to be tolerated as a "binding".

To cater for such wishes, the core of the substitution \(\sigma\) can be
relaxed to contain some passive variables, raising those above the rest,
as it were.
This simple technique is useful beyond the context of renaming,
so we assume arbitrary substitutions.

\begin{definition}[relaxed core]
If \(\Dom(\sigma) \subseteq \{x_1,...,x_n\}\), where variables
\(x_1,...,x_n\) are pairwise distinct, then \(\{x_1,...,x_n\}\) shall be
called a \myemph{relaxed core}
and
\(\left(\mytop{x_1}{\sigma(x_1)} \mytop{...}{...} \mytop{x_n}{\sigma(x_n)}\right) \)
shall be called a \myemph{relaxed core representation}
for \(\sigma\).
If we fix a relaxed core for \(\sigma\), it shall be denoted 
\(\myemphsym{\cplus{\sigma}} \sei \{x_1,...,x_n\}\).
The associated range \(\sigma(\cplus{\sigma})\) we denote as
\(\myemphsym{\rplus{\sigma}}\).
The set of variables of \(\sigma\) is as expected,
\(\myemphsym{\vplus{\sigma}} \sei \cplus{\sigma}\unija\varsof{\rplus{\sigma}}\). 
To get back to the traditional representation, we denote by
\(\myemphsym{\unrelax{\sigma}}\)
the (non-relaxed) core representation of \(\sigma\).
\end{definition}

For extending, substitutions are treated like sets of active bindings,
so (disjoint) union may be used:

\begin{definition}[sum of substitutions]
If
\(\sigma = \left(\mytop{x_1}{s_1} \mytop{...}{...} \mytop{x_n}{s_n}\right)\)  
and
\(\theta = \left(\mytop{y_1}{t_1} \mytop{...}{...} \mytop{y_m}{t_m}\right)\) 
are substitutions in relaxed representation such that
\(\{y_1,..., y_m\} \vardis \{x_1,..., x_n\}\), 
then
\(\myemphsym{\sigma\sadd\theta} \sei 
  \left(\mytop{x_1}{s_1} \mytop{...}{...} \mytop{x_n}{s_n} \mytop{y_1}{t_1} \mytop{...}{...} \mytop{y_m}{t_m}\right)  
\)
is the \myemph{sum} of \(\sigma\) and \(\theta\).
\end{definition}

For \autoref{sec:varlem}, backward compatibility of an extension shall be needed.

\begin{lemma}[backward compatibility]\label{rem:pasiv} 
Let \(\sigma,\theta\) be substitutions and \(x\) be a variable. Then
\((\sigma\sadd\theta)(x)=\sigma(x)\) 
~iff~ \(\theta(x)=x\).
\end{lemma}

\begin{myproof}\reference{rem:pasiv}

If \(x\not\in \cplus{\theta}\), then \(\theta(x)=x\),
and \((\sigma\sadd\theta)(x)=\sigma(x)\).
If \(x\in \cplus{\theta}\), then \((\sigma\sadd\theta)(x)=\theta(x)\)
and also \(x\not\in \cplus{\sigma}\), hence \(\sigma(x)=x\).
The condition \((\sigma\sadd\theta)(x)=\sigma(x)\)
collapses to \(\theta(x)=x\).
\end{myproof}

Passivity of \(\theta\) on a term \(t\) is guaranteed if \(\sigma\) is
"complete" for \(t\), \ie{} lays claim to all its variables:

\begin{definition}[complete for term]
Let \(\sigma\) be given in relaxed core representation. 
We say that \(\sigma\) is
\myemph{complete}
for \(t\) if \(\varsof{t} \subseteq \cplus{\sigma}\).
\end{definition}

In such a case there is no danger that an extension of \(\sigma\) might
map \(t\) differently from \(\sigma\):

\begin{corollary}[backward compatibility]\label{lem:compl:back}
If \(\sigma\) is complete for \(t\), then for any \(\theta\) holds:
\(\sigma\sadd\theta\) is complete for \(t\) and 
\((\sigma\sadd\theta)(t) = \sigma(t)\).
\end{corollary}

\section{\Trenam}\label{sec:trenam}  

In practice, one would like to change the variables in a term without
bothering to check whether this change is a 
permutation of variables, \ie{} a renaming in 
the sense of \autoref{def:ren}. 
For example, the term \(p(z,u,x)\) can be 
changed to 
\(p(y,z,x)\) 
using
mapping
\(z\mapsto y\), \(u\mapsto z\), \(x\mapsto x\).
Let us call such a mapping \emph{\trenam{}}%
\footnote{%
Finding an appropriate name can be a struggle. Shortlisted were
\emph{pre-renaming} and \emph{proto-renaming}.
}.

Like any substitution, a \trenam \(\alpha\) shall also be represented
finitely, but in relaxed core representation, in order to capture
possible \(x\mapsto x\) pairings.  The set \(\cplus{\alpha}\) is fixed
by the terms to map.  Obviously, injectivity is important for such a
mapping, since \(p(z,u,x)\) cannot be mapped on \(p(y,y,x)\) without
losing a variable.  Hence,

\begin{definition}[\trenam]
A \myemph{\trenam} \(\alpha\) is a variable-pure substitution 
injective on a finite set of variables
\({\cplus{\alpha}} \supseteq \Dom(\alpha)\). 
\end{definition}

Clearly, any renaming is a \trenam.  
For \autoref{lem:var}, we need to extend a given \trenam.

\begin{lemma}[extension of \trenam{}]
Let
\(\alpha = \left(\mytop{x_1}{y_1} \mytop{...}{...} \mytop{x_n}{y_n}\right)\)
and
\(\beta = \left(\mytop{u_1}{v_1} \mytop{...}{...} \mytop{u_m}{v_m}\right)\)
be \trenam{}s such that
\(\{u_1,..., u_m\} \vardis \{x_1,..., x_n\}\)  
and
\(\{v_1,..., v_m\} \vardis \{y_1,..., y_n\}\). 
Then \(\alpha\sadd\beta =  
  \left(\mytop{x_1}{y_1} \mytop{...}{...} \mytop{x_n}{y_n} \mytop{u_1}{v_1} \mytop{...}{...} \mytop{u_m}{v_m}\right) 
\)
is also a \trenam, with \(\cplus{\alpha\sadd\beta} = \cplus{\alpha} \dunija \cplus{\beta}\) and
\(\rplus{\alpha\sadd\beta} = \rplus{\alpha} \dunija \rplus{\beta}\).
\end{lemma}

\label{page-transl}
Plotkin's concept of \emph{translation} \cite[p.\,46]{plotkin:phd}
corresponds to \trenam without passive bindings.
There, the \emph{inverse translation} for
\(\tau \sei  \left(\mytop{z}{y} \mytop{u}{z}\right)\)
would be 
\(\tau_{inv} \sei \left(\mytop{y}{z} \mytop{z}{u}\right)\).
Clearly, \(\tau_{inv} \compos \tau =
\left(\mytop{y}{z} \mytop{z}{u}\right) \compos \left(\mytop{z}{y} \mytop{u}{z}\right) 
= \left(\mytop{y}{z}\right)\),
which is not identity substitution.
Although \(\restrict{(\tau_{inv}\compos\tau)}{\{z,u\}} = \idsubst\), 
so \(\tau\) is reversible and thus "safe" to use on \(\{z,u\}\),
one might instinctively be wary of the possibility that handling 
several translations in the same computation could somehow produce "unsafety".
Presumably for that reason, the concept of translation did not catch on,
and it is meanwhile customary to define alphabetic variance 
using renaming rather than translation (\cite{apt:book}).
We revisit Plotkin's concept, enriched with passive bindings
and deemed fit for a new name, \emph{\trenam{}},
and show that its safe application on a term and  
safe (even backward-compatible) extension are easily achievable,
thus justifying the intuitive practice.

\subsection{The question of inverse}

So a \trenam is more natural in practice, but a "full" renaming is 
better mathematically tractable, due to its being invertible on \(\VarSet\).
The next property shows how to extend a \trenam \(\alpha\) to obtain a renaming, 
and a \emph{relevant} one at that, \ie{} 
active only on the variables from \(\vplus{\alpha}\).  
The claim is essentially given in
\cite{lloyd:shepherdson:91}, \cite{apt:book} and \cite{amato:05} 
with emphasis on the core%
\footnote{%
 \cite[p.\,23]{apt:book}: 
"Every finite 1-1 mapping \(f\) from \(A\) onto \(B\) can be extended to
 a permutation \(g\) of \(A\unija B\).  Moreover, if \(f\) has no
 fixpoints, then it can be extended to a \(g\) with no fixpoints."
}
of such an extension.  
Originally the claim appears in \cite{eder:85},
with emphasis on the extent of coincidence%
\footnote{%
\cite[p.\,35]{eder:85}: 
"Let W be a co-finite set of variables (...) and let \(\sigma\) be a 
\hyperlink{bkm-w-ren}{W-renaming}. Then there is a permutation \(\pi\) which 
coincides with  \(\sigma\) on the set W."
},
which is our concern as well.  
We rephrase the claim around the notion of \trenam, 
and provide a constructive proof based on \autoref{lem:cyc}.

\begin{theorem}[embedding]\label{lem:ren:w} 
If \(\alpha\) is a \trenam, there is a renaming 
\(\closure{\alpha}\) 
which coincides with \(\alpha\) on
\(\VarSet\setminus(\rplus{\alpha}\setminus\cplus{\alpha})\) 
such that \(\varsof{\closure{\alpha}}\subseteq \vplus{\alpha}\). 
Additionally, if \(\alpha(x)\not=x\) on \(\cplus{\alpha}\), then
\(\closure{\alpha}(x)\not=x\) on \(\vplus{\alpha}\). 
\end{theorem}

\begin{figure}[htbp]\center
\hrulefill\par
\(\myemphsym{\closure{\alpha}(x)} \sei 
\begin{cases}
\alpha(x), & \text{if } x\in \cplus{\alpha} \\
z, & \text{if } x\in \rplus{\alpha}\setminus \cplus{\alpha} \text{ and } \alpha^{m}(z)=x \text{ for maximal } m\leq n \\
x, & \text{outside of }\cplus{\alpha}\unija \rplus{\alpha}
\end{cases}
\)
\par\hrulefill
\caption{Closure, the natural relevant embedding}  
\label{alg:embed}
\end{figure}

\begin{myproof}\reference{lem:ren:w}
If \(\alpha\) is a \trenam, then \(\cplus{\alpha}\) and \(\rplus{\alpha}\) 
are sets of \(n\) distinct variables each.  
The wanted renaming is constructed in \autoref{alg:embed},
with the intention to close the possibly open chain \(x, \alpha(x), \alpha^2(x),...\)
So let us see whether for every \(x\) there is a \(j\) such that
\(\closure{\alpha}^j(x)=x\).  If \(x\in \cplus{\alpha}\), we start as in the proof of 
\autoref{lem:cyc}, and consider the sequence \(x,\alpha(x),\alpha^2(x), ...\)  
Since \(\cplus{\alpha}\) is finite, either we get two equals (and proceed as there), 
or we get \(\alpha^k(x)\not\in \cplus{\alpha}\) and are stuck.
For \(y\sei \alpha^k(x)\) we know \(\closure{\alpha}(y)=z\) such that
\(\alpha^{m}(z) = y\) with maximal \(m\), so \(m\geq k\).  Therefore,
\(
\alpha^m(\closure{\alpha}(y))=y=\alpha^k(x)
\).
Due to injectivity of \(\alpha\) on \(\cplus{\alpha}\), we get
\(
\alpha^{m-k}(\closure{\alpha}(\alpha^k(x))) = x
\),
and hence \( \closure{\alpha}^{m+1}(x)=x \).

The cases \(x\in \rplus{\alpha}\setminus \cplus{\alpha}\) or \(x\not\in \cplus{\alpha}\unija \rplus{\alpha}\) are easy.
By  \autoref{lem:cyc}, \(\closure{\alpha}\) is injective.
By \autoref{lem:eder:85}, \(\closure{\alpha}\) is a renaming.
The discussion of the case \(\alpha(x)\not= x\) on \(\cplus{\alpha}\) is straightforward.
\end{myproof}

\begin{definition}[closure of a \trenam]
The renaming \(\closure{\alpha}\) constructed in \autoref{alg:embed} shall be called the
\myemph{closure} of \(\alpha\).
\end{definition}
\begin{remark}[relevant embedding is not unique]
Let \(\alpha=\left(\mytop{z}{y} \mytop{u}{z} \mytop{y}{x} \mytop{w_1}{w_2}\right)\),
and let us embed it in a relevant renaming.  
The \autoref{alg:embed} gives
\(\closure{\alpha} = 
  \left(\mytop{z}{y} \mytop{u}{z} \mytop{y}{x} \mytop{w_1}{w_2} \mytop{x}{u} \mytop{w_2}{w_1}\right) 
\). 
But \(\rho = 
  \left(\mytop{z}{y} \mytop{u}{z} \mytop{y}{x} \mytop{w_1}{w_2} \mytop{x}{w_1} \mytop{w_2}{u}\right)  
\)
is also a relevant renaming which is embedding \(\alpha\).  
In the usual notation for cycle decomposition, \(\rho = \{(x,w_1,w_2,u,z,y)\}\)
and \(\closure{\alpha} = \{(x,u,z,y), (w_1, w_2) \}\).
\end{remark}

If we reverse the \trenam, the closure algorithm shall be closing the
same open chains but in the opposite direction, hence
\begin{lemma}[reverse \trenam]\label{lem:cl:inv} 
Let
\(\alpha\sei \left(\mytop{x_1}{y_1} \mytop{...}{...} \mytop{x_n}{y_n}\right)\)  
and
\(\alpha_{inv}\sei \left(\mytop{y_1}{x_1} \mytop{...}{...} \mytop{y_n}{x_n}\right)\). 
Then
\(\closure{\alpha_{inv}}=\closure{\alpha}^{-1}\).
\end{lemma}

\begin{remark}[closure is not compositional]\label{rem:cl:nocomp}
Take
\(\alpha\sei \left(\mytop{z}{y} \mytop{u}{z} \mytop{y}{x}\right)\)
and \(\rho\sei\left(\mytop{x}{y} \mytop{y}{x}\right)\).
Then \(\closure{\alpha}=\left(\mytop{z}{y} \mytop{u}{z} \mytop{y}{x} \mytop{x}{u}\right)\),~
\(\rho\compos\closure{\alpha} = \left(\mytop{z}{x} \mytop{u}{z} \mytop{x}{u}\right)\),~
\(\rho\compos\alpha=\left(\mytop{z}{x} \mytop{u}{z} \mytop{x}{y}\right)\)
and
\(\closure{\rho\compos\alpha}=\left(\mytop{z}{x} \mytop{u}{z} \mytop{x}{y} \mytop{y}{u}\right)\).
\end{remark}

\begin{remark}[closure is not monotone]\label{rem:cl:nomon}
If \(\alpha \supseteq \alpha'\),
then not always 
\(\closure{\alpha} \supseteq \closure{\alpha'}\).
To see this, let
\(\alpha=\left(\mytop{z}{y} \mytop{u}{z} \mytop{y}{x}\right)\) and
\(\alpha'=\left(\mytop{z}{y} \mytop{u}{z}\right)\).
Then
\(\closure{\alpha'}=\left(\mytop{z}{y} \mytop{u}{z} \mytop{y}{u}\right)\)
and
\(\closure{\alpha}=\left(\mytop{z}{y} \mytop{u}{z} \mytop{y}{x} \mytop{x}{u}\right)\).
\end{remark}

\subsection{Staying safe}\label{sec:safe} 

Let us look more closely into \autoref{rem:cl:nomon}: 
\(\alpha(y)=x \text{ and } \alpha(x)=x\), so \(y\) and \(x\) may not
simultaneously occur in the candidate term.  Otherwise, a variable shall
be lost, which we call "aliasing", like in
\(\left(\mytop{y}{x}\right)(p(x,f(y))) = p(x,f(x))\).  

\begin{definition}[aliasing]
Let \(\alpha\) be a \trenam.  If \(x\not= y\) but \(\alpha(x)=\alpha(y)\), then \(\alpha\) is
\myemph{aliasing} \(x\) and \(y\). 
\end{definition}

So what \autoref{rem:cl:nomon} means is: 
if we want to use \(\alpha\) on a larger set than \(\cplus{\alpha}\), then the set
\(\myemphsym{\noninj(\alpha)} \sei \rplus{\alpha}\setminus\cplus{\alpha}\)
should be avoided, because aliasing may happen. 
But, luckily, its complement is safe:

\begin{lemma}[larger set]\label{lem:larger}
A \trenam \(\alpha\) is injective on the co-finite set
\(\VarSet\setminus\noninj(\alpha)\).
The set is maximal containing \(\cplus{\alpha}\).
\end{lemma}

\begin{myproof}\reference{lem:larger}

Let \(x,y \in \VarSet\setminus\noninj(\alpha)\).  Is it possible that
\(\alpha(x)=\alpha(y)\)?  Possible cases: If \(x,y \in \cplus{\alpha}\),
then by definition of \trenam \(\alpha(x)\not=\alpha(y)\).  
If \(x,y \not\in \cplus{\alpha}\), then \(\alpha(x)=x\not= y=\alpha(y)\).  
It remains to consider the mixed case \(x\in\cplus{\alpha}, \,y\not\in \cplus{\alpha}\).  
We have \(\alpha(x) \in \rplus{\alpha}\) and
\(\alpha(y)=y\).  So is \(\alpha(x)=y\) possible?  If yes, then \(y \in \rplus{\alpha}\), but since \(y\not\in \cplus{\alpha}\), that would mean
\(y \in \noninj(\alpha)\).  Contradiction.
 
The set cannot be made larger: if \(y \in \noninj(\alpha)\), then there is 
\(x \in \cplus{\alpha}\) with \(x\not=y\) and \(\alpha(x)=y=\alpha(y)\). 
\end{myproof}

\begin{definition}[injectivity domain]
Since \(\myemphsym{\indom(\alpha)} \sei \VarSet\setminus\noninj(\alpha)\)
is the largest co-finite set containing \(\cplus{\alpha}\)
on which \(\alpha\) is injective, it shall be called
the \myemph{injectivity domain} of \(\alpha\).
\end{definition}

The injectivity domain of a \trenam is clearly the only safe place for it to be mapping terms from.  

\begin{definition}[safety of \trenam]

A \trenam{} \(\alpha\) is \myemph{safe}%
\footnote{%
Safe \trenam is more general than
\emph{renaming for a term} in 
\cite[p.\,22]{lloyd}, since we do not require
\(\Dom(\alpha) \subseteq \varsof{t}\). 
}
for a term $t$ if \(\varsof{t}\subseteq \indom(\alpha)\). 
\end{definition}

Clearly, \(\indom(\alpha) = \cplus{\alpha} \unija (\VarSet\setminus\rplus{\alpha})\), 
so \(\alpha\) is safe for its relaxed core. Hence, 

\begin{corollary}[complete and safe]\label{lem:compl:safe} 
If a \trenam is complete for a term, it is safe for that term.
\end{corollary}

For a \trenam \(\alpha\) with the quality \(
\rplus{\alpha}=\cplus{\alpha} \), \ie{} a renaming, it is no surprise
that \(\indom(\alpha)=\VarSet\) and hence safety is guaranteed for any term.

A \trenam behaves like a renaming on its injectivity domain, since it
coincides with its closure there.  This follows immediately from \autoref{lem:ren:w}:
\begin{lemma}[injectivity domain]\label{lem:indom}  
Let \(x\in \indom(\alpha)\). Then \(\alpha(x) = \closure{\alpha}(x)\).
\end{lemma}

\begin{corollary}[\trenam stability]\label{lem:ren:aff:gen}  
A generalization of \autoref{lem:ren:aff} holds:
Let \(s,t\) be terms and \(\alpha\) be a \trenam safe for \(s,t\).  
Then \(s=t\) iff 
\(\alpha(s)=\alpha(t)\) and also \(s\in t\) iff \(\alpha(s) \in \alpha(t)\).  
As a consequence, \(s\vardis t\) iff \(\alpha(s) \vardis \alpha(t)\).
\end{corollary}

Our definition of \trenam was inspired by the following more general notion from \cite{eder:85}.
\label{page-w-ren}
\begin{definition}[W-renaming, \cite{eder:85}]
Let \(W\subseteq V\).  A substitution \(\sigma\) is a \myemph{\hypertarget{bkm-w-ren}{W-renaming}}  
if \(\sigma\) is variable-pure on \(W\), and \(\sigma\) is injective on \(W\).
\end{definition}
With this notion, \autoref{lem:larger} can be summarized as:
\(\indom(\alpha)\) is a co-finite set of variables,
and the largest set \(W \supseteq \cplus{\alpha}\) such that \(\alpha\)
is a W-renaming.

What about safety of extension?  If \(\alpha\) is safe for \(t\),
\(\alpha\sadd\beta\) does not have to be, even if \(\beta(t)=t\), as the
following example shows: 
\( \alpha\sei\left(\mytop{v}{w}\right),~ \beta\sei\left(\mytop{z}{y} \mytop{u}{z} \mytop{y}{x}\right),~ t\sei p(x) \)
(here no aliasing happened, though). 
The next two claims address safety of extension.

\begin{lemma}[monotonicity]\label{lem:cont}
Assume \(\alpha\sadd\beta\) is defined. Then
\begin{enumerate}
\item
\(\indom(\alpha) \unija\, \indom(\beta) = \VarSet\)
\item
\(\indom(\alpha) \presek\, \indom(\beta)  \subseteq \indom(\alpha\sadd\beta)\)
\end{enumerate}
\end{lemma}

\begin{myproof}\reference{lem:cont}
Since
\(
(\VarSet\setminus A)\unija (\VarSet\setminus B) = \VarSet\setminus (A\presek B))
\),
and \(\noninj(\alpha) \vardis \noninj(\beta)\), we get
\(
\indom(\alpha) \unija \indom(\beta) = \VarSet
\).
Further, 
\(
(\VarSet\setminus A) \presek (\VarSet\setminus B) = \VarSet\setminus (A\unija B)
\)
and so
\(
\noninj(\alpha\sadd\beta) = (\rplus{\alpha}\dunija\rplus{\beta}) 
  \setminus (\cplus{\alpha}\dunija\cplus{\beta}) 
  \subseteq 
   (\rplus{\alpha}\setminus\cplus{\alpha}) \unija (\rplus{\beta}\setminus\cplus{\beta})
  = \noninj(\alpha) \unija \noninj(\beta)
\).
\end{myproof}

In \autoref{rem:cl:nomon}, \(\noninj(\alpha') = \{y\}\), 
\(\noninj(\left(\mytop{y}{x}\right)) = \{x\}\),  
and \(\noninj(\alpha) = \{x\}\),
hence
\(\indom(\alpha') = \VarSet\setminus\{y\}\),
\(\indom(\left(\mytop{y}{x}\right)) = \VarSet\setminus\{x\}\)  
and
\(\indom(\alpha) = \VarSet\setminus\{x\}\).

By the last claim, staying within \(\indom(\alpha)\) and
\(\indom(\beta)\) ensures staying within \(\indom(\alpha\sadd\beta)\).
By assuming a bit more about \(\alpha\) than just safety, we may
ignore the nature of extension \(\beta\), and still ensure safety and
even backward compatibility of \(\alpha\sadd\beta\).  
This shall be used in \autoref{sec:appl}.

\begin{theorem}[safety of extension]\label{lem:exx}
Assume \(\alpha\sadd\beta\) is defined.
\begin{enumerate}
\item
If \(\alpha\) is safe for \(t\) \,and\, \(\beta\) is safe for \(t\),
then \(\alpha\sadd\beta\) is safe for \(t\).

\item \label{lemx:eq}
If \(\alpha\) is complete for \(t\), then
\(\alpha\sadd\beta\) is complete (hence safe) for \(t\), and
\((\alpha\sadd\beta)(t) = \alpha(t)\).
\end{enumerate}
\end{theorem}
\noindent
The first part follows from \autoref{lem:cont} and the second
from \autoref{lem:compl:back} and \autoref{lem:compl:safe}.

\subsection{Variant of term and substitution}

The traditional notion of term variance, which is term renaming, shall
be generalized to \trenam.  As a special case, substitution variance is
defined, inspired by substitution renaming from \cite{amato:05}.  
For this, substitution shall once again be regarded as a special case of
term.  The term is of course the relaxed core representation.  This
concept shall come in handy for proving properties of renamed
derivations (\autoref{sec:varlem}).

\subsubsection{Term variant}

\begin{definition}[term variant]
If \(\alpha\) is a \trenam safe for \(t\), then $\alpha(t)$ is a 
\myemph{variant} of \(t\),  written 
\myemphsym{\(\alpha(t) \variant t\)}. 
The particular variance and the direction of its application may be
explicated by 
\myemphsym{\(s \variant[\alpha] t\)} iff \(s=\alpha(t)\). 
\end{definition}

If \(s\variant t\), then there is a unique \(\alpha\) mapping \(s\) to
\(t\) in a complete and \cumulative%
\footnote{%
"\Cumulative{}" in the binary sense (page \pageref{page-cumul}). 
In case of \trenam, we naturally use \cplus{} as \realDom and \(\rplus{}\) as \realRange. 
}
manner, \ie{} mapping each variable pair and nothing else, as computed
by \autoref{alg:pren}.  The algorithm makes do with only one set for
equations and bindings, thanks to different types.
Termination can be seen from the tuple
\((\mathit{lfun}_{=}(E), \mathit{card}_{=}(E))\) decreasing in
lexicographic order with each rule application,
where \(\mathit{lfun}_{=}(E)\) is the number of function symbols in
equations in \(E\), and \(\mathit{card}_{=}(E)\) is the number of equations in \(E\).

\begin{figure}[htbp]
\hrulefill\par%
Start from the set \(E \sei \{s=t\}\) and transform according to the
following rules.  The transformation is bound to stop.  If the stop was
not due to failure, then the final set \(E\) is 
{the \trenam of \(s\) to \(t\)}, \myemphsym{\(\pren(s,t)\)}.
\begin{description}
\item[elimination]
\(
E \dunija \{x=y\}  \goto E, \text{ if } x/y \in E
\)
\item[failure: alias]
\(
E \dunija \{x=y\} \goto \text{failure, if }(x/z\in E,\, z\not=y)\text{ or } (z/y\in E,\, z\not=x)  
\)
\item[binding]
\(
E \dunija \{x=y\} \goto E\unija\{x/y\}, \text{ if }(x/\any\not\in E)\text{ and } (\any/y\not\in E)   
\)
\item[failure: instance]
\(
E \dunija \{x=t\}  \goto \text{failure, if } t\not\in \VarSet
\);~
\(
E \dunija \{t=x\}  \goto \text{failure, if } t\not\in \VarSet
\)
\item[decomposition]
\(
E \dunija \{f(s_1,...,s_n)=f(t_1,...,t_n)\} \goto E\unija\{s_1=t_1,..., s_n=t_n\}  
\)
\item[failure: clash]
\(
E \dunija \{f(s_1,...,s_n)=g(t_1,...,t_m)\} \goto \text{failure, if }f\not=g\text{ or }m\not=n 
\)
\end{description}
\unskip
\hrulefill
\caption{Computing the \trenam of $s$ to $t$}  
\label{alg:pren}
\end{figure}

\begin{notation}[epsoid]
The \trenam constructed in \autoref{alg:pren} shall be simply called
\myemph{the \trenam} of \emph{\(s\) to \(t\)}, 
and denoted \myemphsym{\(\pren(s,t)\)}. 
It is complete for \(s\) and \cumulative for \(s\) to \(t\).

In case \(s=t\), we obtain for \(\pren(s,t)\) essentially the identity
substitution.  However, regarded as \trenam{}s, \(\pren(t,t)\) and
\idsubst are not the same.  A \trenam \(\alpha\) with relaxed core \(W\)
mapping each variable on itself (in other words, \(\cplus{\alpha} = W\)
and \(\unrelax{\alpha}=\idsubst\)) shall be called the \myemph{$W$-epsoid} 
and denoted \(\myemphsym{\id{W}}\).
For a term \(t\), we abbreviate
\(\myemphsym{\id{t}} \sei \id{\varsof{t}}\).
\end{notation}
Regarding composition, an epsoid behaves just like \idsubst.  
Its use is for providing completeness, and hence extensibility,
by means of placeholder bindings \(x/x\).

\subsubsection{Special case: substitution variant}\label{sec:substvar}
Even substitutions themselves can be renamed.  To rename a
substitution, one regards it as a syntactical object, a set of
bindings, and renames those bindings.
If \(\rho\) is a renaming and \(\sigma\) is a substitution,
\cite{amato:05} defines substitution renaming by
\( 
\rho(\sigma) \sei \{ \rho(x)/\rho(\sigma(x)) ~|~ x \in \Dom(\sigma) \}
\). 
It is easy to see that \(\rho(\sigma)\) is a substitution in core
representation.  For this only two properties of \(\rho\) were needed:
variable-purity on \(\varsof{\sigma}\) and injectivity on
\(\varsof{\sigma}\).  These requirements are clearly fulfilled by
\trenam{}s safe on \(\sigma\) as well.  Hence,

\begin{definition}[substitution variant]\label{def:proto} 
Let \(\sigma\) be a substitution and let \(\alpha\) be a
\trenam safe for \(\sigma\),  \ie{} 
\(\varsof{\sigma}\subseteq \indom(\alpha)\). 
Then a
\myemph{variant} \emph{of $\sigma$ by $\alpha$} is
\begin{equation}\label{eq:rho:sigma}
\myemphsym{\alpha(\sigma)} \sei \{ \alpha(x)/\alpha(\sigma(x)) ~|~ x \in \Dom(\sigma) \}
\end{equation}
\end{definition}
We may write 
\myemphsym{\(\theta \variant[\alpha] \sigma\)} 
if \(\theta = \alpha(\sigma)\), as with any other terms.  As can be
expected, the concept of variance by \trenam is well-defined, owing to
safety.  Otherwise, the result of \trenam would not even have to be a
substitution again, as in the case of
\(\alpha=\left(\mytop{y}{x}\right)\) and \(\sigma=\left(\mytop{x}{a} \mytop{y}{b}\right)\).

\begin{lemma}[well-defined]\label{lem:welldef}
Substitution variant is well-defined, \ie{} \eqref{eq:rho:sigma} 
is a core representation of a substitution, and \(\alpha\) does not introduce aliasing. 
\end{lemma}

\begin{myproof}\reference{lem:welldef}

Let \(\Dom(\sigma)=\{x_1,...,x_n\}\).
Due to injectivity of \(\alpha\) on \(\varsof{\sigma}\), if
\(\alpha(x_i)=\alpha(x_j)\), then \(x_i=x_j\), so \(i=j\).  
To finish the proof that \eqref{eq:rho:sigma} a core representation, observe
\( x \in \Dom(\sigma) \) iff \( x \not= \sigma(x) \) iff 
\( \alpha(x) \not= \alpha(\sigma(x)) \), due to injectivity again.
Re aliasing, by \autoref{lem:ren:aff:gen}, if
\(\alpha(\sigma(x_i)) \nonvardis \alpha(\sigma(x_j))\),
then \(\sigma(x_i) \nonvardis \sigma(x_j)\),
meaning that \(\alpha\) does not introduce aliasing.
\end{myproof}

From \autoref{def:proto} and \autoref{lem:indom} follows 
\begin{lemma}
Let \(\sigma\) be a substitution, \(\alpha, \beta\) be \trenam{}s and 
\(\alpha(\sigma)\) and \((\alpha\compos\beta)(\sigma)\) be defined.
Then
\begin{enumerate}
\item
\( (\alpha\compos\beta)(\sigma) = \alpha(\beta(\sigma)) \)

\item
\( \alpha(\sigma) = \closure{\alpha}(\sigma) \)
\end{enumerate}
\end{lemma}

For the case of "full" renaming, there is a way to dissolve the new expression:%
\footnote{%
an immediate consequence being \(\rho(\sigma)\neq\rho\compos\sigma\) 
} 
\begin{legacy}[\cite{amato:05}]\label{lem:amato}
For any renaming \(\rho\) and substitution \(\sigma\)
\[
\rho(\sigma) = \rho\compos\sigma\compos\rho^{-1}
\]
\end{legacy}
Would such a claim hold for the weakened case, \trenam{}s?

\begin{theorem}[substitution variant]\label{lem:proto} 
Let \(\sigma\) be a substitution and \(\alpha\) be a \trenam safe for
\(\sigma\). Then
\begin{enumerate}
\item
\(\alpha(\sigma)\compos\alpha = \alpha\compos\sigma\)
\item
\(\alpha(\sigma)=\closure{\alpha}\compos\sigma\compos\closure{\alpha}^{-1}\)
\end{enumerate}
\end{theorem}

\begin{myproof}\reference{lem:proto}

First part:
According to \autoref{def:proto}, for every \(x\in\VarSet\) holds 
\( (\alpha(\sigma)\compos\alpha)(x) = \alpha(\sigma(x)) \).  
Since any substitution is structure-preserving, the claim holds for any term \(t\) as well.
Second part: From the first part we know
\(\closure{\alpha}(\sigma)\compos\closure{\alpha}=\closure{\alpha}\compos\sigma\),
hence
\(\closure{\alpha}(\sigma)=\closure{\alpha}\compos\sigma\compos\closure{\alpha}^{-1}\).
By \autoref{lem:indom}, \( \alpha(\sigma) = \closure{\alpha}(\sigma) \).  
\end{myproof}

It is known that idempotence and equivalence of substitutions are not
compatible with composition \cite{eder:85}.  Luckily, the concept of
variance, with constant \trenam, does not share this handicap:

\begin{theorem}[compositionality]\label{lem:proto:compos} 
Let \(\sigma,\theta\) be substitutions and \(\alpha\) be their safe \trenam. Then
\begin{align*}
\alpha(\sigma\compos\theta) &= \alpha(\sigma)\compos\alpha(\theta) 
\end{align*}
\end{theorem}

\begin{myproof}\reference{lem:proto:compos}

Since \(\varsof{\sigma\compos\theta} \subseteq \varsof{\sigma}\unija \varsof{\theta}\),
clearly \(\varsof{\sigma\compos\theta} \subseteq  \indom(\alpha)\).
By \autoref{lem:proto},
\(
\alpha(\sigma) \compos \alpha(\theta)  
= \closure{\alpha}\compos\sigma\compos\closure{\alpha}^{-1}\compos\closure{\alpha}\compos\theta\compos\closure{\alpha}^{-1}
= \closure{\alpha}\compos\sigma\compos\theta\compos\closure{\alpha}^{-1}
= \alpha(\sigma\compos\theta)   
\).
\end{myproof}

\section{Application}\label{sec:appl}
Implementing logic programming means that the {freedom} of Horn clause logic 
(\myemphsym{HCL}) must be restrained:
\begin{itemize}
\item
most general unifier (\myemph{mgu}) is provided by a fixed algorithm,

\item
standardization-apart is provided by a fixed algorithm.
\end{itemize}
Every implementation of HCL is parametrized by the two algorithms.
Here we shall consider only the unification algorithm, so by
\myemphsym{\hclparam} 
an implementation of HCL using unification algorithm \ualg is denoted.
From the literature (\emph{variant lemma}) we know that such a restriction is not
compromising soundness and completeness of SLD-resolution. Yet, there may
be lots of "lowlier" claims which more or less implicitly rely on
freedom of mgus and standardization-apart.  For example, with both
choices fixed we may not any more just rename an SLD-derivation
wholesale (the resolvents, the mgus, the input clauses), as was possible
in Horn clause logic, based on \autoref{lem:ren:aff}.  This is because
the two algorithms do not have to be renaming-compatible.  In fact, the
second one cannot be, which makes claims like \autoref{lem:propag} necessary.

Let us cast a look at the first restriction.  For any two unifiable
terms \(s,t\) holds that the set of their mgus, written as
\(\myemphsym{\mgus(s,t)}\), is infinite.  
On the other hand, in practice any unification algorithm \ualg produces, 
for the given two unifiable terms, just one deterministic value as their mgu.  
We shall denote this particular mgu of \(s\) and \(t\) as
\(\myemphsym{\ualg(s, t)}\), 
the \myemph{algorithmic mgu} of \(s\) and \(t\) obtained by \ualg.

The abundancy of mgus is not only good, it also stands in the way of proofs.  
The simplest unification problem \(p(x)=p(y)\)
has among others two equally attractive candidate mgus,
\((\mytop{x}{y})\) and \((\mytop{y}{x})\). 
Assume our unification algorithm decided upon \((\mytop{x}{y})\). 
Assume further that we rename the protagonists and obtain \(p(x)=p(z)\).
What mgu shall be chosen this time?  To ensure some dependability in
this issue, we shall require of any unification algorithm the following
simple requirement, postulated as an axiom:

\begin{axiom}[renaming compatibility]\label{ax:mgu} 
Let \ualg be a unification algorithm.
For any renaming \(\rho\) and any equation \(E\), it has to hold
\(
\ualg(\rho(E)) = \rho(\ualg(E)).
\)
\end{axiom}
Since classical unification algorithms like Robinson's and
Martelli-Montanari's do not depend upon the actual names of variables
(as observed in \cite{amato:05}), 
this requirement is in practice always satisfied.

\begin{remark}[renaming compatibility of \mgus]\label{rem:mgu}
For every \(\rho\) and \(E\) holds
\(
\mgus(\rho(E)) = \rho(\mgus(E)).
\)
This is due to \autoref{lem:proto} and \autoref{lem:ren:aff}.
Assume \(\sigma\in\mgus(s,t)\), then
\(
\rho(\sigma)(\rho(s))= \rho(\sigma(s))= \rho(\sigma(t)) = \rho(\sigma)(\rho(t))
\). 
Further, if \(\theta\) is a unifier of \(\rho(s),\rho(t)\),
then \(\theta\compos\rho\) is a unifier of \(s,t\),
hence there is a renaming \(\delta\) with \(\theta\compos\rho = \delta\compos\sigma\), giving 
\(\theta
=\delta\compos\sigma\compos\rho^{-1}
=\delta\compos\rho^{-1}\compos\rho\compos\sigma\compos\rho^{-1} = (\delta\compos\rho^{-1})\compos\rho(\sigma)\),
meaning \(\rho(\sigma)\in \mgus(\rho(E))\).
For the other direction, observe
\(\theta 
= \rho\compos\rho^{-1}\compos\delta\compos\sigma\compos\rho^{-1}
= \rho(\rho^{-1}\compos\delta\compos\sigma)
\). 
\end{remark}

\subsection{Handling local variables in \protect\hclparam}\label{sec:local} 

With \ualg complying to \autoref{ax:mgu} and yielding relevant mgus, that is to say 
with practically any \ualg,%
\footnote{Classical unification algorithms not only satisfy \autoref{ax:mgu} but also 
yield idempotent mgus. Idempotent mgus are always relevant (\cite{apt:book}).
}
\label{foot-relevance}
a propagation result for SLD-derivations can be proved, which leads to a
constructive and incremental version of the variant lemma.

Regarding SLD-derivations, for the most part we shall assume traditional
concepts as given in \cite{lloyd} and \cite{apt:book}, but with some changes and
additions listed below.
An input clause \(\clausi_i\) obtained from a program clause \(\klausi\)
by replacing the variables in order of appearance with \(t_1,...,t_n\)
may be denoted as \(\clausi_i=\myemphsym{\klausi[t_1,...,t_n]}\).
Assume now an SLD-derivation \compd for \(G\) of the form
\(
G \resol{\clausi_1\dopp\sigma_1} G_1 
\resol{\clausi_2\dopp\sigma_2}
...
\resol{\clausi_n\dopp\sigma_n}
G_n
\).

\begin{itemize}
\item

$\clausi_i$ is here the \emph{actually used} variant of a program clause 
(\ie, the current \emph{input clause}) and not the program clause itself.

\item
The substitution \(\sigma_n\compos...\compos\sigma_1\) shall be called 
the \myemph{partial answer} for \(G\) at step \(n\) of the derivation.
A final partial answer, whenever \(G_n=\emptycl\), shall be called a
\myemph{\casq} for \(G\).

\end{itemize}
Relation to earlier concepts of answer: In a derivation, the \myemph{resultant} 
of level \(n\) is defined in \cite{lloyd:shepherdson:91} as
\(
\sigma_n\compos...\compos\sigma_1(G) \gneck  G_n
\).  
So partial answer is at heart of resultant.
An \emph{answer substitution} 
for \(G\) is defined in \cite{apt:vanemden:82} exactly as \casq; 
later (\cite{lloyd}, \cite{apt:book})
it is made relevant by restricting to variables of \(G\), 
hence \(\restrict{(\sigma_n\compos...\compos\sigma_1)}{G}\) whenever \(G_n=\emptycl\),
and called \myemph{computed answer substitution (\cas)}. 

Showing the actually used variants of program clauses (instead of
program clauses themselves) enables a simple definition of derivation variables: 
the annotations
\(\clausi_i\dopp\sigma_i\) 
are regarded as part of the derivation, so
\(
\myemph{\varsof{\compd}}  \sei 
(\varsof{G} \unija ... \unija \varsof{G_n})
\unija
(\varsof{\sigma_1} \unija ... \unija \varsof{\sigma_n})
\unija
(\varsof{\clausi_1} \unija ... \unija \varsof{\clausi_n})
\).

Now to the propagation result. Assume the program 
"\meta{son(S) \gneck male(S),\, child(S,P).}",  
and let us enquire about \meta{son} in two derivations. 
If we know that one query, say \meta{son(X)}, is a variant of the
other, \meta{son(A)}, does the same connection hold between the
resolvents as well?

As can be seen from \autoref{tab:resolve} and \autoref{fig:propag}, 
in a resolution some new variables may crop up, originating from standardization-apart.
Let us call them \myemph{local variable}s,  as opposed to \emph{query variable}s.
Likely causes are
a pattern in a clause head (\fsig[1]{nat} in \autoref{tab:resolve})
or surplus variables in a clause body (\fsig[1]{son} in \autoref{fig:propag}). 
But even without those, 
local variables can appear (\fsig[1]{p} in \autoref{tab:resolve}), 
except with a restriction to \emph{normal SLD-derivation} (\cite{bol:92}), 
preventing "needless renaming of variables". 
Were it not for local variables, the resolvents in both derivations
would clearly be variants, with the same \trenam as for the original queries.
Yet, even though the variables new in one derivation do not have to be new in the other, 
the \trenam can (under reasonable conditions) be extended to accomodate them.
The claim is proved in a constructive manner.

\begin{table}[htbp]\center
\begin{tabular}{@{}l@{\hspace{2\tabcolsep}}l@{\hspace{2\tabcolsep}}l@{}}
\begin{cell}
\rm query
\\\midrule
   \meta{nat(X)} 
\\
  \meta{nat(A)} 
\\
 \meta{p(X)}
\\
 \meta{p(A)} 
\end{cell}
&
\begin{cell} 
\rm  input clause
\\\midrule
\meta{nat(s(A)) \gneck nat(A).} 
\\
\meta{nat(s(B)) \gneck nat(B).} 
\\
\meta{p(A) \gneck q(A).} 
\\
\meta{p(B) \gneck q(B).} 
\end{cell}
&
\begin{cell} 
\rm resolvent
\\\midrule
\meta{nat(A)}   
\\
\meta{nat(B)} 
\\
\meta{q(X)} or \meta{q(A)}
\\
\meta{q(A)} or \meta{q(B)}
\end{cell}
\end{tabular}
\caption{Resolution may produce local variables} 
\label{tab:resolve}
\end{table}

\begin{lemma}[propagation of variance]\label{lem:propag}
Assume a unification algorithm \ualg satisfying \autoref{ax:mgu}.
Assume an SLD-derivation \(\compd\) ending with \(G\) and an
SLD-derivation \(\compd'\) ending with \(G'\) such that
\(\alpha(G)=G'\) for some \trenam \(\alpha\)
which is complete for \(G\) and \cumulative for 
\(\compd\) to \(\compd'\).

Further assume that
 \(G \resol{\clausi\dopp\sigma} H\)
and
 \(G' \resol{\clausi'\dopp\sigma'} H'\)
such that in \(G\) and \(G'\) atoms in the same positions were selected
and \(\clausi, \clausi'\) are variants.
Lastly assume that \(\sigma\) is a relevant mgu.
Then for \(\lambda \sei \pren(\clausi, \clausi')\) holds 
\begin{enumerate}

\item\label{lem:propag:complete}
\(\alpha\sadd\lambda\) is complete for \(H\)  
and \(\sigma\)

\item\label{lem:propag:cumulative}
\(\alpha\sadd\lambda\)  
is \cumulative for 
\(\compd \resol{\clausi\dopp\sigma} H\) to
\(\compd' \resol{\clausi'\dopp\sigma'} H'\) 

\item\label{lem:propag:increm}
\(H' = (\alpha\sadd\lambda)(H)\)
and
\(\sigma' = (\alpha\sadd\lambda)(\sigma)\)

\end{enumerate}
\end{lemma}
\begin{figure}[htbp]\center
\begin{math}
\begin{CD}
G @>\resol{}>>  H\\ 
@V\alpha V\alpha\,\sadd\,\lambda V   
@VV\alpha\,\sadd\,\lambda V\\  
G'  @>>\resol{}>  H'    
\end{CD}
\end{math}
\hspace*{2cm} 
\begin{math}
\begin{CD}
\meta{son(X)} @>\resol{}>> \meta{male(X),\, child(X,B)} \\ 
@VV\alpha=\left(\mytop{X}{A} \mytop{B}{X}\right)V 
@VV\alpha\,\sadd\,\text{?} V\\  
\meta{son(A)}  @>>\resol{}>  \meta{male(A),\, child(A,C)} 
\end{CD}
\end{math}
\caption{Propagation of variance ...is not always possible}
\label{fig:propag}
\end{figure}

The claim can be summarized in \autoref{fig:propag}, together with the r\^{o}le of \cumulativeness requirement. 
From \autoref{lem:propag} and \autoref{tab:resolve}
follows that resolution is not \trenam-stable (or even renaming-stable);
extending the \trenam with local variables may be necessary. 

\begin{myproof}\reference{lem:propag}

First let us establish that \(\alpha\sadd\lambda\) is defined.
Due to 
\emph{\cumulativeness of \(\alpha\) for \(\compd,\compd'\)},
\begin{equation}\label{lemp:cumul0}
{\cplus{\alpha}} \subseteq \varsof{\compd} \text{~and~}  {\rplus{\alpha}} \subseteq \varsof{\compd'}  
\end{equation}
Due to \emph{standardization-apart},
\(\clausi \vardis {\compd}\)  and \(\clausi' \vardis {\compd'}\), hence
\begin{align}
&{\cplus{\lambda}} \vardis {\compd}  \text{~and~} {\rplus{\lambda}} \vardis {\compd'} \label{lemp:stap}  
\end{align}
Thus
\( {\cplus{\alpha}} \vardis {\cplus{\lambda}} \) and \( {\rplus{\alpha}} \vardis {\rplus{\lambda}} \),
so \(\alpha\sadd\lambda\) is defined.
Also,
\eqref{lemp:stap} proves that \(\lambda\) is passive on old variables, \ie{}
\(\lambda(\compd)=\compd\). 
Let \(G \sei (M,A,N)\) and \(H \sei \sigma(M,B,N)\), 
where \(M,B,N\) may be conjunctions. 
Then \( G' = (M',A',N')=(\alpha(M),\alpha(A),\alpha(N)) \).
Let \(\clausi\from A_1 \gneck B_1\) and
\(\clausi'\from A_2 \gneck B_2\). Then
\(\sigma = \ualg(A, A_1)\), \(B = \sigma(B_1)\) 
and \(\sigma' = \ualg(A', A_2)\), \(B' = \sigma'(B_2)\). 
Also, 
\begin{align}
& \alpha \text{ is complete for }M,A,N\text{: } \varsof{(M,A,N)} \subseteq \cplus{\alpha}, \text{ by \emph{completeness} of \(\alpha\) for \(G\)} \label{lemp:ac} 
\\
& \lambda \text{ is complete for }A_1,B_1\text{: } \varsof{\clausi} = \varsof{(A_1, B_1)} = \cplus{\lambda}, \text{ by definition of \(\lambda\)} \label{lemp:lc} 
\end{align}
Due to \emph{relevance} of \(\sigma\), we have
\(\varsof{\sigma} \subseteq \varsof{A}\unija\varsof{A_1}\),
which together with \eqref{lemp:ac} and \eqref{lemp:lc} gives
\begin{align}
& \alpha\sadd\lambda \text{ is complete for }\sigma\text{: } \varsof{\sigma} \subseteq 
\cplus{\alpha\sadd\lambda} 
\label{lemp:cs}
\end{align}
Having thus fielded all the assumptions, we obtain
\begin{align}
& \alpha\sadd\lambda \text{ is safe for }M,A,N,A_1, B_1,\sigma\text{: } \text{ by \eqref{lemp:ac}, \eqref{lemp:lc}, \eqref{lemp:cs} and \autoref{lem:exx}\eqref{lemx:eq}}  \label{lemp:safe}  
\end{align}
\noindent
{\em Proof of \ref{lem:propag:complete}.:}
Completeness of \(\alpha\sadd\lambda\) for \(\sigma\) is proved above. 
Completeness of \(\alpha\sadd\lambda\) for \(H\) follows from
\(
\varsof{H} \subseteq \varsof{G} \unija \varsof{\clausi}
   \subseteq \cplus{\alpha} \unija \cplus{\lambda}
   = \cplus{\alpha\sadd\lambda}
\),
by \eqref{lemp:ac} and \eqref{lemp:lc}.

\medskip
\noindent
{\em Proof of \ref{lem:propag:cumulative}.:}
By definition, \(\cplus{\lambda} = \varsof{\clausi}\)
and \(\rplus{\lambda} = \varsof{\clausi'}\). 
Hence, and due to \cumulativeness of \(\alpha\),
\(  
\cplus{\alpha\sadd\lambda} = \cplus{\alpha} \dunija \cplus{\lambda}
\subseteq \varsof{\compd} \unija \varsof{\clausi}  
\subseteq \varsof{\compd \resol{\clausi\dopp\sigma} H}.
\)  
Similarly,
\(\rplus{\alpha\sadd\lambda} \subseteq \varsof{\compd' \resol{\clausi'\dopp\sigma'} H'}\),
therefore 
\(\alpha\sadd\lambda\) is \cumulative for
\(\compd \resol{\clausi\dopp\sigma} H\)
to
\(\compd' \resol{\clausi'\dopp\sigma'} H'\).

\medskip
\noindent
{\em Proof of \ref{lem:propag:increm}.:}
\begin{align*}
\lefteqn{\sigma' = \ualg(A',A_2)  = \ualg(\alpha(A), \lambda(A_1)) 
=\ualg((\alpha\sadd\lambda)(A), (\alpha\sadd\lambda)(A_1)), \text{ by \eqref{lemp:ac}, \eqref{lemp:lc} and \autoref{lem:compl:back}}  }
\\&
  = \ualg((\closure{\alpha\sadd\lambda})(A), (\closure{\alpha\sadd\lambda})(A_1))
= \closure{(\alpha\sadd\lambda)}(\sigma) = {(\alpha\sadd\lambda)}(\sigma) , \text{ by \eqref{lemp:safe}, \autoref{lem:indom} and \autoref{ax:mgu}}
\displaybreak[0]\\[.4em] 
\lefteqn{B' = \sigma'(B_2) = (\alpha\sadd\lambda)(\sigma)(\lambda(B_1))    
=(\alpha\sadd\lambda)(\sigma)((\alpha\sadd\lambda)(B_1)), \text{ by \eqref{lemp:lc} and \autoref{lem:compl:back}}  }
\\&  
=(\alpha\sadd\lambda)(\sigma(B_1))=(\alpha\sadd\lambda)(B), \text{ by \autoref{lem:proto}}
\displaybreak[0]\\[.4em]
\lefteqn{H' = \sigma'(\alpha(M), B', \alpha(N)) 
= (\alpha\sadd\lambda)(\sigma)(\alpha(M), (\alpha\sadd\lambda)(B), \alpha(N)), \text{ by preceding line}   }
\\&
= (\alpha\sadd\lambda)(\sigma)((\alpha\sadd\lambda)(M), (\alpha\sadd\lambda)(B), (\alpha\sadd\lambda)(N)), \text{ by \eqref{lemp:ac} and \autoref{lem:compl:back}} 
\\&
= (\alpha\sadd\lambda)(\sigma(M,B,N)) 
= (\alpha\sadd\lambda)(H), \text{ by \autoref{lem:proto}}
\end{align*}  
\end{myproof}

\subsection{Variant lemma for \protect\hclparam}\label{sec:varlem}

\begin{definition}[similarity]
SLD-derivations of the same length
\begin{equation}\label{eq:def:simil}
\begin{split}
G \resol{\clausi_1\dopp\sigma_1} G_1 \resol{\clausi_2\dopp\sigma_2}  ... \resol{\clausi_{n}\dopp\sigma_{n}} G_{n}
\\
G' \resol{\clausi'_1\dopp\sigma'_1} G'_1 \resol{\clausi'_2\dopp\sigma'_2}  ... \resol{\clausi'_{n}\dopp\sigma'_{n}} G'_{n} 
\end{split}
\end{equation}
are
\myemph{similar}
if \(G\) and \(G'\) are variants and additionally at each step \(i\)
holds: atoms in the same position are selected, and the input clauses
\(\clausi_i\) and \(\clausi'_i\) are variants. 
\end{definition}

That the name "similarity" is justified, follows from the claim known as
\emph{variant lemma} (\cite{lloyd}, \cite{lloyd:shepherdson:91}, \cite{apt:book}),
here in the formulation from \cite{doets:93}:
\emph{Finite derivations which are similar and start from variant queries have variant resultants}.
For logic programming systems obeying \autoref{ax:mgu} and
relevance of mgu, a more precise claim can be proved.  The added
assumptions (the axiom and relevance) are practically void (see footnote
on page \pageref{foot-relevance}), yet the added conclusion has
substance: first, renaming a query now costs a degree of freedom -- if we
treat the two variants of the program clause at each step as independent, 
then the two mgus are not independent.  
Second, the precise variance is now known.
 
\begin{theorem}[variant claim for \hclparam]\label{lem:var}
Assume a unification algorithm \ualg satisfying \autoref{ax:mgu}
and yielding relevant mgus. Then:
\begin{itemize}
\item
finite SLD-derivations which are similar and start from variant queries
have variant partial answers

\item
the variance depends only on the starting queries and input clauses.
\end{itemize}
In particular, assume our similar derivations to be as in
\eqref{eq:def:simil}.  Then for every \(i=1,...,n\) holds
\(
G_i' = \beta_i(G_i),~ 
\sigma_i' = \beta_i(\sigma_i)
\text{~and~} \sigma_i'\compos...\compos\sigma_1' = \beta_i(\sigma_i\compos...\compos\sigma_1)
\),
where
\(\beta_i \sei \alpha\sadd\lambda_1\sadd...\sadd\lambda_i\),~
\(\alpha \sei \pren(G, G')\)~
and
\(\lambda_i \sei \pren(\clausi_i, \clausi_i')\).
\end{theorem}

\begin{myproof}\reference{lem:var}
By assumption, \(G\) and \(G'\) are variants, so
\begin{equation}\label{lemv:1}
\alpha \sei \pren(G, G')
\end{equation}
exists.
Clearly, \(\alpha\) is complete for \(G\). 
By construction, \(\alpha\) is also
\cumulative for \(\compd_0 \sei G\) ~to~ \(\compd_0' \sei G'\).
We may iterate \autoref{lem:propag}, obtaining for every \(i=1,...,n\)
\begin{align}
\sigma_i' &= (\alpha\sadd\lambda_1\sadd...\sadd\lambda_i)(\sigma_i) \label{lemv:2}
\\
G_i' &= (\alpha\sadd\lambda_1\sadd...\sadd\lambda_i)(G_i) \label{lemv:3}
\end{align}
where \(\lambda_i \sei \pren(\clausi_i, \clausi_i')\).
Due to completeness of \(\alpha\) for \(G\) and \autoref{lem:compl:back},
\begin{align} \label{lemv:4a}
\alpha(G) &= (\alpha\sadd\lambda_1\sadd...\sadd\lambda_i)(G)
\end{align}
So
\(
\sigma_i'\compos \sigma_{i-1}'\compos...\compos\sigma_1' =
  (\alpha\sadd\lambda_1\sadd...\sadd\lambda_i)(\sigma_i) 
  \compos 
  (\alpha\sadd\lambda_1\sadd...\sadd\lambda_{i-1})(\sigma_{i-1})
  \compos 
...
  \compos
  (\alpha\sadd\lambda_1)(\sigma_1)
\).
We would like to extract \(\sigma_i\compos
\sigma_{i-1}\compos...\compos\sigma_1\) on the right, to have a connection between partial answers.

Assume
\(k<i\). Since \(\varsof{\sigma_k} \subseteq \varsof{G} \unija \varsof{\clausi_1}\unija...\unija \varsof{\clausi_k} 
\subseteq \cplus{\alpha}\unija\cplus{\lambda_1}\unija...\unija\cplus{\lambda_k}
= \cplus{\alpha\sadd\lambda_1\sadd...\sadd\lambda_k}
\),
by \autoref{lem:exx}\eqref{lemx:eq}
\(
(\alpha\sadd\lambda_1\sadd...\sadd\lambda_k\sadd...\sadd\lambda_i)(\sigma_k) =
(\alpha\sadd\lambda_1\sadd...\sadd\lambda_k)(\sigma_k) 
\).
Hence,
\begin{equation}
\begin{split}\label{lemv:4}
  (\alpha\sadd\lambda_1\sadd...\sadd\lambda_{i-1})(\sigma_{i-1}) &= (\alpha\sadd\lambda_1\sadd...\sadd\lambda_i)(\sigma_{i-1})
\\
...
\\
  (\alpha\sadd\lambda_1)(\sigma_1) &= (\alpha\sadd\lambda_1\sadd...\sadd\lambda_i)(\sigma_1) 
\end{split}
\end{equation}
Let us abbreviate
\(
\beta_i \sei \alpha\sadd\lambda_1\sadd...\sadd\lambda_i
\).
Then from \eqref{lemv:2} and \eqref{lemv:4} by
\autoref{lem:proto:compos}
\begin{equation}\label{lemv:5}
\sigma_i'\compos \sigma_{i-1}'\compos...\compos\sigma_1' =
  \beta_i(\sigma_i) \compos \beta_i(\sigma_{i-1})\compos...\compos \beta_i(\sigma_1)
= \beta_i(\sigma_i\compos\sigma_{i-1}\compos...\compos\sigma_1)
\end{equation}
which is the promised connection between partial answers.

Clearly, variance of partial answers means variance of \casq{}s, 
\cas{}es and resultants as well. 
For the cases when \(G_n = \emptycl\), we obtain, by \eqref{lemv:5},
the expected relationship between the respective \casq{}s:
\(\sigma_n'\compos...\compos\sigma_1' = \beta_n(\sigma_n\compos...\compos\sigma_1)
\).
\Cas{} differs from \casq{} by the added restriction on the query variables.
Due to renaming-compatibility of restriction, \eqref{lemv:5} and
\(\beta_n(G) = \alpha(G) = G'\), we obtain
\(
\restrict{\sigma_n'\compos...\compos\sigma_1'}{G'} = \beta_n(\restrict{\sigma_n\compos...\compos\sigma_1}{G}) 
\), 
\ie{} the same relationship.
Finally, knowing that the {resultant} of step \(i\) is
\({R_i} \sei \sigma_i\compos...\compos\sigma_1(G) \gneck G_i\),
we obtain
\begin{align*}
\lefteqn{R_i' = 
  \sigma_i'\compos...\compos\sigma_1'(G') \gneck G_i'
  = \beta_i(\sigma_i\compos...\compos\sigma_1)(\alpha(G)) \gneck \beta_i(G_i), \text{ by \eqref{lemv:5}, \eqref{lemv:1} and \eqref{lemv:3}}  }
\\&
= \beta_i(\sigma_i\compos...\compos\sigma_1)(\beta_i(G)) \gneck \beta_i(G_i)  
= \beta_i((\sigma_i\compos...\compos\sigma_1)(G)) \gneck \beta_i(G_i), \text{ by \eqref{lemv:4a} and \autoref{lem:proto}} 
\\& 
= \beta_i(R_i)  
\end{align*}
So partial answers of step \(i\) (and in consequence \cas{}es and resultants) 
are variant via \(\beta_i\).
\end{myproof}

\begin{example}[similarity]
Let the logic program be
\begin{alignat*}{2}
&son(X) \gneck male(X),\, child(X,A). && \quad\color{\Commentcolor}\%\, \klausi_1
\\
&male(c).~ male(d).~ child(d,a).    && \quad\color{\Commentcolor}\%\, \klausi_2,~ \klausi_3, ~\klausi_4
\end{alignat*}
An interpreter for LD-resolution may produce the following two derivations:
\begin{align*}
&
\meta{son(A)} \resol{\clausi_1\dopp\sigma_1} \meta{male(A),\, child(A,C)} \resol{\clausi_2\dopp\sigma_2} \meta{child(d,C)}  
\\&
\meta{son(B)} \resol{\clausi'_1\dopp\sigma'_1} \meta{male(B),\, child(B,D)} \resol{\clausi'_2\dopp\sigma'_2} \meta{child(d,D)}  
\end{align*}
They are obviously similar, with
\(\clausi_1 = \klausi_1[\specu,C]\),~
\(\clausi'_1 = \klausi_1[\specv,D]\),~
\(\clausi_2 = \clausi'_2 = \klausi_3\). 
The variables \(\specu,\specv\) stand for actually used variables, which are not
discernible from the form of derivations.
From the queries, input clauses and resolvents we can further deduce
which relevant mgus were used: 
\(\sigma_1 = \left(\mytop{\specu}{A}\right)\),~
\(\sigma'_1 = \left(\mytop{\specv}{B}\right)\),~
\(\sigma_2 = \left(\mytop{A}{d}\right)\) and \(\sigma'_2 = \left(\mytop{B}{d}\right)\). 
The mappings are
\(\alpha=\left(\mytop{A}{B}\right)\),~
\(\lambda_1=\left(\mytop{\specu}{\specv} \mytop{C}{D}\right)\) and \(\lambda_2=\idsubst\).
Clearly, they fulfill
\((\alpha\sadd\lambda_1)(\meta{male(A),\, child(A,C)}) = \meta{male(B),\, child(B,D)}\)  
and
\((\alpha\sadd\lambda_1)(\left(\mytop{B}{A}\right))= \left(\mytop{C}{B}\right)\),  
as well as
\((\alpha\sadd\lambda_1\sadd\lambda_2)(\meta{child(d,C)})=\meta{child(d,D)}\),  
and so forth. 
\end{example}

\section{Outlook} 

Concepts relating to variable renaming have been reviewed and built upon,
with the aim of providing for practical needs of program analysis and
formal semantics in logic programming.  
By \emph{relaxing the core representation} and forgoing permutation
requirement for renaming, the concept of \emph{\trenam{}} is obtained.
It is a mathematical underpinning of the intuitive practice of
renaming terms by just considering the necessary bindings,
where now \(x/x\) may be necessary.
In other words, a \trenam is a variable-pure substitution with 
mutually distinct variables in range, possibly including some passive bindings.

\Trenam{}s enable incremental claims like variance propagation
in implemented Horn clause logic (\autoref{lem:propag}, \autoref{lem:var}).
There, \trenam{}s made it possible to keep track of new (\emph{local}) variables.

\end{document}